\documentclass[fleqn,10pt]{wlscirep}
\usepackage{lineno}
\usepackage{color}
\usepackage{comment}
\usepackage{subfigure}
\usepackage{graphicx}
\usepackage{multirow}
\usepackage{siunitx}
\usepackage{cite}
\usepackage{caption}
\usepackage{threeparttable}
\usepackage{tablefootnote}
\usepackage{algorithm, algorithmic}
\usepackage{xspace}
\usepackage{booktabs}
\usepackage{marvosym}
\usepackage{ifsym}
\usepackage{makecell}
\usepackage{newtxtext,newtxmath}
\usepackage{bm}
\usepackage{amsfonts,amssymb} 
\usepackage{colortbl}
\usepackage{xcolor}
\usepackage{tikz}
\newcommand*{\circled}[1]{\lower.7ex\hbox{\tikz\draw (0pt, 0pt)%
    circle (.45em) node {\makebox[1em][c]{\small#1}};}}

\usepackage{amsfonts, bm}

\def\eqref#1{equation~\ref{#1}}

\def\1{\bm{1}}

\DeclareMathAlphabet{\mathsfit}{\encodingdefault}{\sfdefault}{m}{sl}
\SetMathAlphabet{\mathsfit}{bold}{\encodingdefault}{\sfdefault}{bx}{n}

\title{
    \begin{minipage}{0.1\textwidth}
        \centering
        \includegraphics[height=2.0cm]{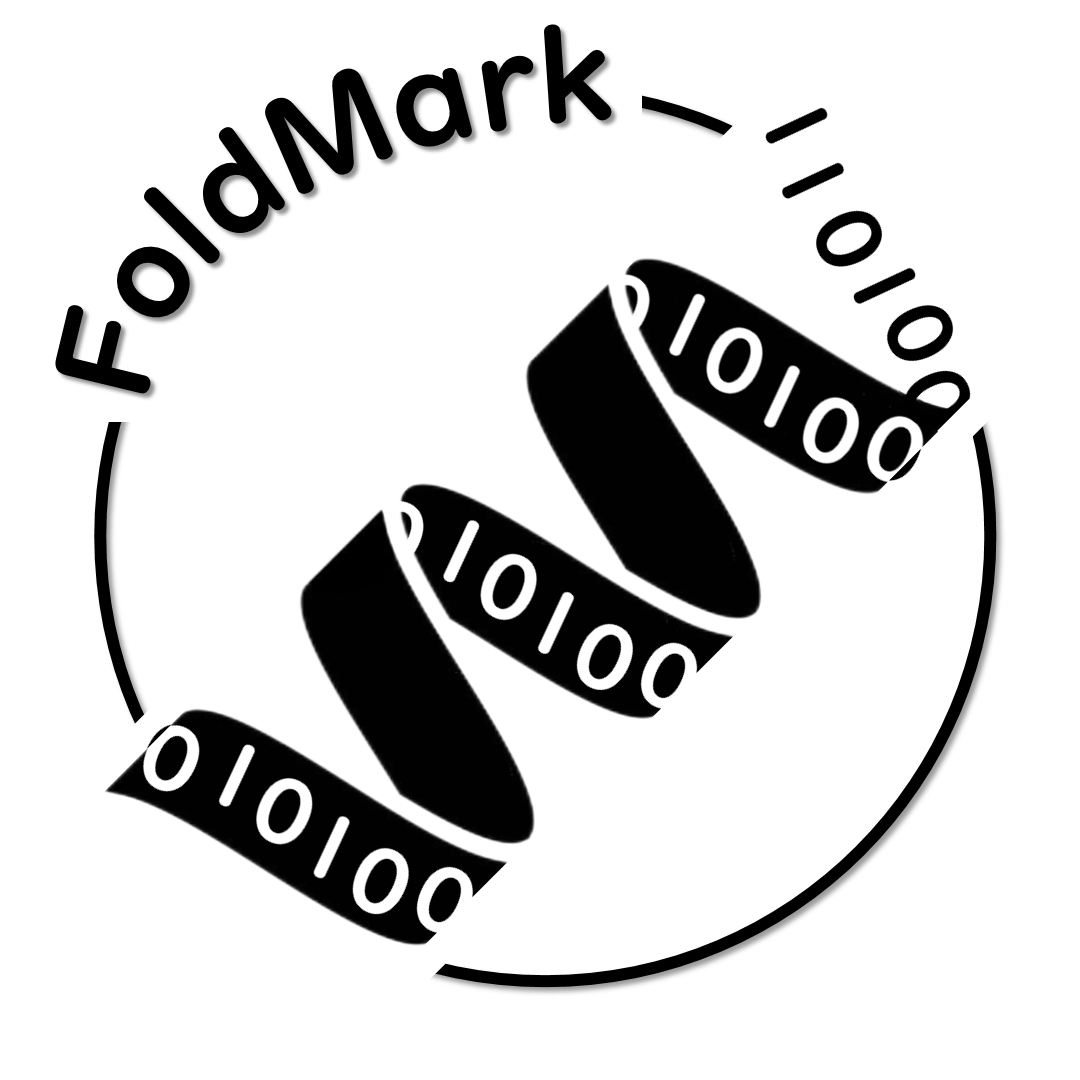} 
    \end{minipage}
    \hspace{0.02\textwidth}
    \begin{minipage}{0.81\textwidth}
        \centering
        FoldMark: Protecting Protein Generative Models with Watermarking
    \end{minipage}
}

\author[1]{\centering Zaixi Zhang}
\author[2]{\centering Ruofan Jin}
\author[3]{\centering Kaidi Fu}
\author[4]{\centering Le Cong}
\author[5]{\centering Marinka Zitnik}
\author[1,\Letter]{Mengdi Wang}
\affil[1]{Princeton University, NJ, USA}
\affil[2]{Zhejiang University, Zhejiang, China}
\affil[3]{Tsinghua University, Beijing, China}
\affil[4]{Stanford University, CA, USA}
\affil[5]{Department of Biomedical Informatics, Harvard Medical School, Boston, MA, USA}

\affil[\Letter]{mengdiw@princeton.edu}

\begin{abstract}
Protein structure is key to understanding protein function and is essential for progress in bioengineering, drug discovery, and molecular biology. Recently, with the incorporation of generative AI, the power and accuracy of computational protein structure prediction/design have been improved significantly. However, ethical concerns such as copyright protection and harmful content generation (biosecurity) pose challenges to the wide implementation of protein generative models. 
Here, we investigate whether it is possible to embed watermarks into protein generative models and their outputs for copyright authentication and the tracking of generated structures.  
As a proof of concept, we propose a two-stage method FoldMark as a generalized watermarking strategy for protein generative models. 
FoldMark first pretrain watermark encoder and decoder, which can minorly adjust protein structures to embed user-specific information and faithfully recover the information from the encoded structure. In the second step, protein generative models are fine-tuned with Watermark-conditioned Low-Rank Adaptation (WaterLoRA) modules to preserve generation quality while learning to generate watermarked structures with high recovery rates.
Extensive experiments are conducted on open-source protein structure prediction models (e.g., ESMFold and MultiFlow) and de novo structure design models (e.g., FrameDiff and FoldFlow) and we demonstrate that our method is effective across all these generative models. 
Meanwhile, our watermarking framework only exerts a negligible impact
on the original protein structure quality and is robust under potential post-processing and adaptive attacks. 
 
\end{abstract}
\begin{document}

\flushbottom
\maketitle
%
%
\thispagestyle{empty}
\section*{Introduction}
Proteins are life’s essential building blocks and the basis of all living organisms. Understanding their structure is key to uncovering the mechanisms behind their function. 
With the advancement of generative AI \cite{stokel2023chatgpt}, protein structure generative models have revolutionized both protein structure prediction \cite{jumper2021highly, abramson2024accurate} and de novo protein design \cite{watson2023novo, ingraham2023illuminating}, opening up a wide range of applications in bioengineering and drug discovery. 
For example, AlphaFold2 \cite{jumper2021highly} made a breakthrough by accurately predicting protein structures from amino acid sequences at near-experimental accuracy, solving a decades-old challenge in biology. Its successor, AlphaFold3 \cite{abramson2024accurate}, further improved on this by enhancing the ability to model more complex protein interactions and assemblies. Meanwhile, RFDiffusion \cite{watson2023novo} and Chroma \cite{ingraham2023illuminating} introduced diffusion-based generative models that enable the creation of novel protein structures with desired properties and functions, pushing the boundaries of de novo protein design.
In recognition of the profound impact of these models, the 2024 Nobel Prize in Chemistry was awarded to David Baker ``for computational protein design" and to Demis Hassabis and John M. Jumper ``for protein structure prediction" \cite{nobel2024chemistry}.

However, the rapid development of protein generative models and the lack of corresponding regulations lead to copyright and biosecurity concerns. 
First, the ease of model sharing brings up copyright concerns, including the risk of unauthorized use of generated structures and the redistribution of pretrained models for profit, which could undermine the interests of the original creators \cite{jo2023promise, epstein2023art, mesko2023imperative}. For example, the latest AlphaFold3 is only deployed on the server with the terms saying that users ``must not use AlphaFold Server or its outputs to train machine learning models" or ``in connection with any commercial activities''\cite{alphafoldterms2024}.
Second, the unregularized yet powerful protein generative models are vulnerable to misuse and cause bio-security/safety concerns \cite{bloomfield2024ai, baker2024protein}. For example, protein generative models can be used to design new proteins with harmful properties, such as pathogens \cite{wang2022evasion}, toxins \cite{jurenas2022biology}, or viruses \cite{tan2020sars} that can be used as bioweapons. Therefore, there is an urgent need for a reliable and efficient tool to track and audit the use of protein generative models.

Similar problems have occurred in text and image generation. For example, Large language models (LLMs) such as ChatGPT can be used to create fake news and to cheat on academic writing \cite{bergman2022guiding, mitchell2023detectgpt, wu2024fake}. The latest text-to-image models such as Stable Diffusion \cite{rombach2022high} and DALL·E 3 \cite{betker2023improving} enable users to create photo-realistic images like deep fakes \cite{westerlund2019emergence} for illegal purposes. 
As a result, there is growing consensus that the ability to detect, track, and audit the use of AI-generated content is essential for harm reduction and regulation \cite{mitchell2023detectgpt, zhang2022deepfake}. Recently, the watermark becomes one of the most promising protection strategy, which embeds hidden patterns in
the generated content and is imperceptible to humans, while making the embedded information algorithmically identifiable. Although watermark has been applied for LLM \cite{kirchenbauer2023watermark, liu2023unforgeable, zhang2024remark, liu2024survey, Dathathri2024} (e.g., SynthID-Text \cite{Dathathri2024}) and text-to-image models \cite{fernandez2023stable, min2024watermark, feng2024aqualora, yang2024gaussian} (e.g., AquaLoRA \cite{feng2024aqualora}), extending these methods to protein structure data presents unique challenges. Unlike text and images, protein structures are highly sensitive to minute changes \cite{jumper2021highly,liu2024novo, van2024fast}, and embedding watermarks without disrupting the biological functionality or stability of the proteins is a complex task. Moreover, protein structures exhibit complex geometrical symmetries, making traditional watermarking methods less effective due to the requirement for equivariance \cite{watson2023novo, lu2024dynamicbind}.

In this paper, as a proof of concept, we propose FoldMark, a generalized watermarking method for protein generative models, e.g., AlphaFold2 \cite{jumper2021highly} and RFDiffusion \cite{watson2023novo}. FoldMark builds upon pretrained protein generative models and generally has two training stages. In the first stage, SE(3)-equivariant watermark encoder and decoder are pretrained to learn how to embed watermark information into protein structure without compromising the structural quality. Specifically, the encoder takes the watermark code and learns to construct the watermark-conditioned structure; the decoder takes the watermarked structure and predicts the embedded watermark. In the second stage, we propose WaterLoRA \cite{hu2021lora}, which flexibly encodes the given watermark code and merges it into the original model weights, without changing or adding extra model architecture. 
The protein generative model is fine-tuned with two objectives, named message retrieval loss and consistency loss. 
The message retrieval loss ensures the effective
embedding of watermarks into the generated structure, allowing for successful retrieval of the embedded watermark codes. Meanwhile, the consistency loss ensures that
the inclusion of the watermark has minimal impact on the overall quality of the protein structure.

In experiments, we first observe that FoldMark achieves nearly 100\% bit accuracy on watermark code recovery from encoded protein structures with minimal influence on structural validity (measured by scRMSD and RMSD), which means FoldMark can reliably embed and retrieve watermarks with minimal structural deviation. Compared with baseline methods, FoldMark achieves consistent improvements and can successfully handle 16-bit watermark code. 
We further consider two application scenarios: detection for copyright protection and identification for tracing the creator. Specifically, in detection, FoldMark aims to verify whether a structure is generated by a certain model; in identification, FoldMark aims to find the exact user that generates the structure by watermark matching (see Fig. \ref{illustration}). Finally, we demonstrate the robustness of FoldMark under multiple post-processing (e.g., Cropping, Rotation+Translation, Coordinates Noising) and adaptive attacks (Finetuning attack and Multi-Message attack).
Collectively, FoldMark efficiently embeds watermarks while maintaining the quality of protein structure data, offering a new approach to ensuring biosecurity in protein design in the era of generative AI.

\section*{Results}
\subsection*{Overview of FoldMark}
\begin{figure*}[ht]
	\centering
  \includegraphics[width=0.98\linewidth]{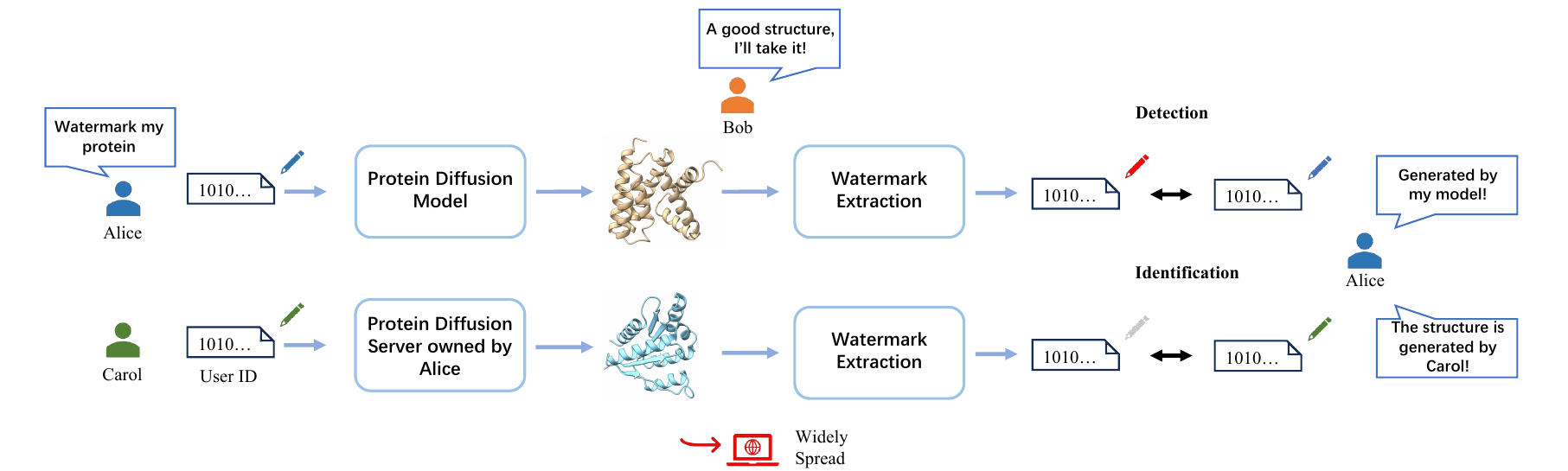}
	\caption{Illustrations of application scenarios of FoldMark. The scenario involves the model owner Alice, the thief Bob, and the user Carol. Alice is responsible for training the model, releasing the pretrained model and inference code, or deploying it on the platform for users. Bob, who downloads Alice's model and code, generates protein structures and falsely claims ownership of the copyrights. Carol registers as a user on the server and utilizes the API to generate protein structures. Alice, as the model owner, may wish to restrict the use of these generated structures, particularly in commercial contexts, to avoid copyright infringement. Using FoldMark, Alice can embed a watermark within the generated protein structures, allowing her to extract the watermark code for detection and identification of unauthorized usage, safeguarding her intellectual property.}
	\label{illustration}
\end{figure*}

\begin{figure*}[ht]
	\centering
  \includegraphics[width=0.98\linewidth]{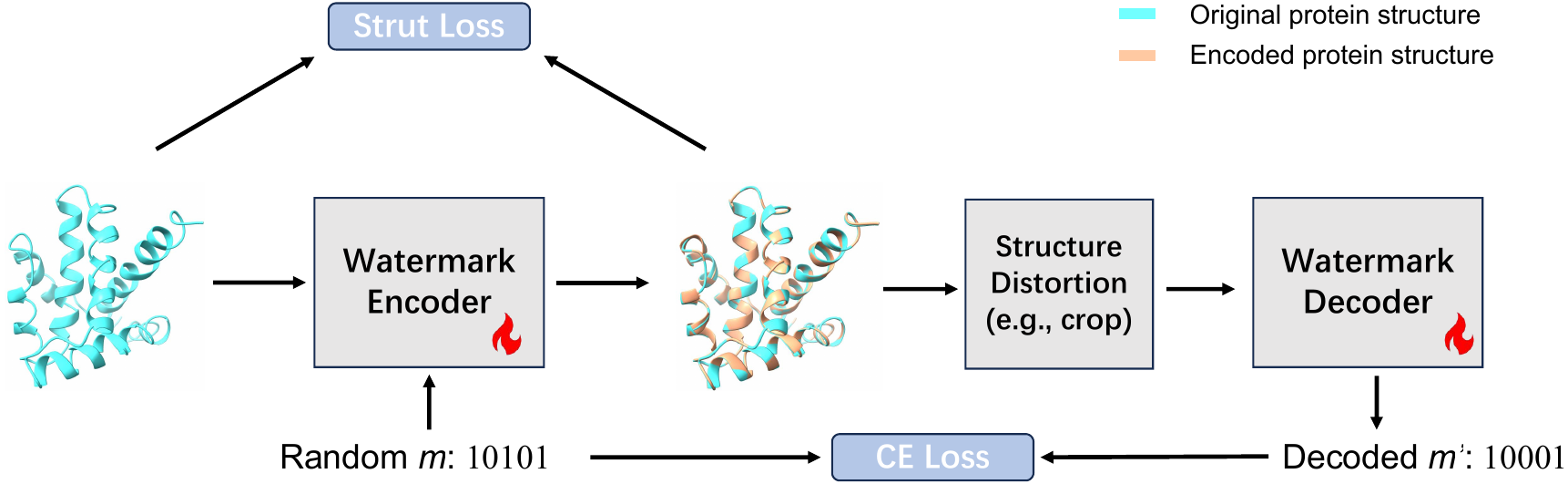}
	\caption{Pretraining stage of FoldMark.}
	\label{pretrain}
\end{figure*}
\vspace{20em}
\begin{figure*}[ht]
	\centering
  \includegraphics[width=0.98\linewidth]{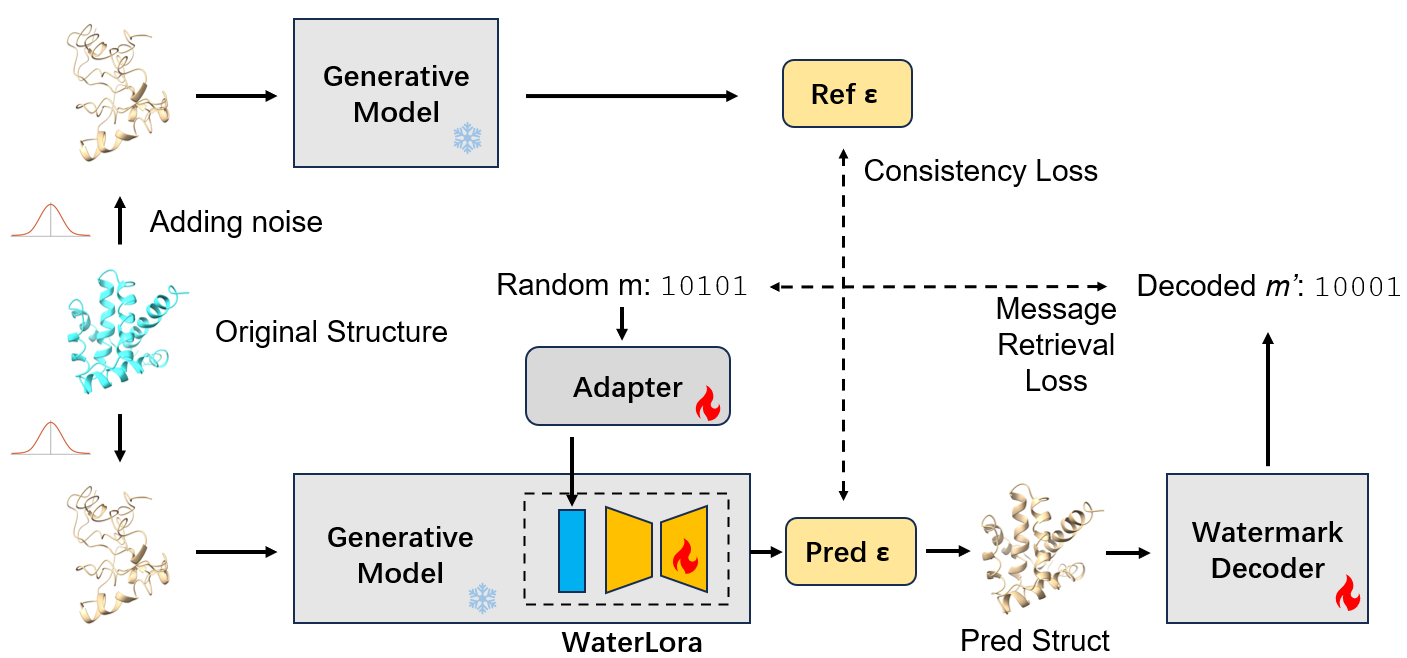}
	\caption{Finetuning stage of FoldMark.}
	\label{finetune}
\end{figure*}

\subsection*{Watermarking State-of-the-art protein generative models}
In Table. \ref{tab:unconditional} and \ref{tab:prediction}, we showed experiments of watermarking unconditional protein structure generative models (i.e., FoldFlow \cite{bose2024se3stochastic}, FrameDiff \cite{yim2023se}, and FrameFlow \cite{yim2023fast}) and protein structure prediction models (i.e., MultiFlow \cite{campbell2024generative} and ESMFold \cite{lin2023evolutionary}). We vary the watermark code length from 4 to 32 and measure the bit prediction accuracy (BitAcc) and the structural validity (scRMSD and RMSD). To benchmark the performance of FoldMark, we also adapt two watermark methods from the image domain, WaDiff \cite{min2024watermark} and AquaLoRA \cite{feng2024aqualora} for comparison. Generally, the performance degrades with the increase of watermarking capacity, i.e., more watermark bits. On most cases with less than 16 bits, FoldMark achieves nearly 100\% bit accuracy on watermark code recovery from encoded protein structures with minimal influence on structural validity (measured by scRMSD and RMSD). Therefore, FoldMark is a generalized and effective method for protein generative model protection. 
\subsection*{Applications in detection and user identification}
As shown in Figure \ref{illustration}, we show two applications of FoldMark. The scenario involves Alice, the model owner responsible for training, releasing the pretrained model, and deploying the inference code on the platform. Bob, a thief, downloads Alice's model and code to generate protein structures, falsely claiming ownership of the copyrights. Carol registers as a user on the server and utilizes the API to generate protein structures. In the detection, the successful extraction of a watermark from structures serves as proof of Alice’s rightful ownership of the copyright and indicates that the structure is artificially generated. In user identification, Alice assigns a unique watermark to each user. By extracting the watermark from generated structures, it becomes possible to trace it back to Carol by comparing it with the watermark database and regarding the most similar user id. Traceability goes beyond detection, enabling copyright protection for different users by identifying the source of infringement. In Table \ref{tab:identification}, we show the identification accuracy for different generative models with different numbers of users. While FoldMark achieves strong performance with small groups of users, it becomes much more challenging for identification among a larger number of users (e.g., $10^6$).
\begin{table*}[htbp]
\caption{Watermarking unconditional protein structure generative models}
\label{tab:unconditional}
\centering
\begin{small}
\begin{tabular}{c|c|cc|cc|cc|cc}
\toprule
\multicolumn{2}{c|}{\multirow{2}{*}{\textbf{Watermark Method}}} & \multicolumn{2}{c|}{\textbf{4bit}} & \multicolumn{2}{c|}{\textbf{8bit}} &\multicolumn{2}{c|}{\textbf{16bit}}&\multicolumn{2}{c}{\textbf{32bit}}\\ 
\cmidrule(lr){3-4} \cmidrule(lr){5-6} \cmidrule(lr){7-8} \cmidrule(lr){9-10}
\multicolumn{2}{c|}{} & BitAcc $\uparrow$ & scRMSD $\downarrow$ & BitAcc $\uparrow$ & scRMSD $\downarrow$ & BitAcc $\uparrow$ & scRMSD $\downarrow$ & BitAcc $\uparrow$ & scRMSD $\downarrow$\\ 
\midrule
\multirow{4}{*}{\makecell[c]{FoldFlow}} & No watermark & - & 1.926 & - & 1.926 & - & 1.926 & - & 1.926 \\
& WaDiff & 88.2\% & 2.107 & 85.0\% & 2.115 & 80.1\% & 2.397 & 64.3\% & 2.630 \\
& AquaLoRA & 73.5\% & 2.056 & 72.6\% & 2.210 & 71.7\% & 2.446 & 62.8\% & 2.718 \\
& FoldMark(Ours) & \textbf{99.9\%}  & \textbf{1.937} & \textbf{99.7\%} & \textbf{1.980}  & \textbf{98.9\%} & \textbf{2.114} &\textbf{94.5\%} & \textbf{2.307}\\ \midrule
\multirow{4}{*}{\makecell[c]{FrameDiff}} & No watermark & - & 2.850 & - & 2.850 & - & 2.850 & - & 2.850 \\
& WaDiff & 76.8\% & 2.919 & 73.3\% & 3.235 & 62.2\% & 3.810 & 50.4\% & 4.058 \\
& AquaLoRA & 64.3\% & 3.150 & 59.1\% & 3.431 & 56.2\% & 3.890 & 51.6\% & 4.179 \\
& FoldMark(Ours) & \textbf{98.7\%} & \textbf{2.795} & \textbf{98.3\%} & \textbf{2.914}  & \textbf{88.4\%} &\textbf{3.045} & \textbf{82.0\%}&\textbf{3.428} \\ \midrule
\multirow{4}{*}{\makecell[c]{FrameFlow}} & No watermark & - & 1.855 & - & 1.855 & - & 1.855 & - & 1.855 \\
& WaDiff & 77.1\% & 1.883 & 76.4\% & 2.270 & 63.5\% & 2.456 & 54.6\% & 2.823 \\
& AquaLoRA & 63.6\% & 1.920 & 61.4\% & 2.317 & 54.5\% & 2.680 & 52.1\% & 2.953 \\
& FoldMark(Ours) & \textbf{99.6\%} & \textbf{1.860} & \textbf{99.5\%} & \textbf{1.939}  & \textbf{96.7\%} &\textbf{2.019} & \textbf{95.4\%}&\textbf{2.192}\\
\bottomrule
\end{tabular}
\end{small}
\end{table*}

\begin{table*}[htbp]
\caption{Watermarking protein structure prediction models}
\label{tab:prediction}
\centering
\begin{small}
\begin{tabular}{c|c|cc|cc|cc|cc}
\toprule
\multicolumn{2}{c|}{\multirow{2}{*}{\textbf{Watermark Method}}} & \multicolumn{2}{c|}{\textbf{4bit}} & \multicolumn{2}{c|}{\textbf{8bit}} &\multicolumn{2}{c|}{\textbf{16bit}}&\multicolumn{2}{c}{\textbf{32bit}}\\ 
\cmidrule(lr){3-4} \cmidrule(lr){5-6} \cmidrule(lr){7-8} \cmidrule(lr){9-10}
\multicolumn{2}{c|}{} & BitAcc $\uparrow$ & RMSD $\downarrow$ & BitAcc $\uparrow$ & RMSD $\downarrow$ & BitAcc $\uparrow$ & RMSD $\downarrow$ & BitAcc $\uparrow$ & RMSD $\downarrow$\\ 
\midrule
\multirow{4}{*}{\makecell[c]{MultiFlow}} & No watermark & - & 3.344 & - & 3.344 & - & 3.344 & - & 3.344 \\
& WaDiff & 64.3\% & 4.342 & 62.0\% & 4.583 & 56.4\% & 5.746 & 52.1\% & 5.643 \\
& AquaLoRA & 65.0\% & 3.626 & 58.9\% & 4.010 & 51.2\% & 4.398 & 50.4\% & 5.277 \\
& FoldMark(Ours) & \textbf{99.7\%}  & \textbf{3.520} & \textbf{98.6\%} & \textbf{3.534}  & \textbf{97.0\%} & \textbf{3.570} &\textbf{89.5\%} & \textbf{3.688}\\ \midrule
\multirow{4}{*}{\makecell[c]{ESMFold}} & No watermark & - & 2.241 & - & 2.241 & - & 2.241 & - & 2.241 \\
& WaDiff & 79.5\% & \textbf{2.426} & 77.1\% & 2.643 & 72.6\% & 2.779 & 59.0\% & 2.800 \\
& AquaLoRA & 76.4\% & 2.446 & 71.8\% & 2.547 & 66.3\% & 2.605 & 54.4\% & 2.820 \\
& FoldMark(Ours) & \textbf{94.6\%} & 2.453 & \textbf{93.2\%} & \textbf{2.506}  & \textbf{86.9\%} &\textbf{2.571} & \textbf{85.0\%}&\textbf{2.680} \\
\bottomrule
\end{tabular}
\end{small}
\end{table*}

\begin{table*}[ht]
\caption{Performance of FoldMark under post-processing and adaptive attacks. Protein post-processing include structure cropping (keeping 50\% of the whole sequence), randomly translating \& rotating the whole structure, and adding Gaussian noise to the coordinates (strength 0.2). Adaptive attacks include fine-tuning the watermarked model with clean protein data to erase the watermarking capability and multi-message attack that try to inject additional watermarks to cover the original ones. We conduct experiments on the 16bits setting.}
\centering
\label{tab:identification}
\begin{tabular}{lcccccc}
\toprule
\textbf{Model} & \textbf{No Attack} & \textbf{Cropping} & \textbf{Trans\&Rotate} & \textbf{Noising} & \textbf{Finetune} & \textbf{Multi-Msg} \\
\midrule
FoldFlow  & 0.989 & 0.961 & 0.990 & 0.910 & 0.920 & 0.947  \\
FrameDiff  & 0.884 & 0.860 & 0.882 & 0.793 & 0.769 & 0.860  \\
FrameFlow  & 0.967 & 0.906 & 0.960 & 0.871 & 0.870 & 0.948  \\
MultiFlow  & 0.970 & 0.864 & 0.972 & 0.826 & 0.924 & 0.950 \\
ESMFold & 0.869 & 0.829 & 0.874 & 0.805 & 0.856 & 0.862 \\
\bottomrule
\end{tabular}
\end{table*}

\begin{table*}[ht]
\caption{Performance of FoldMark user identification accuracy.}
\centering
\label{tab:adaptive}
\begin{tabular}{lcccc}
\toprule
\textbf{Model} & \textbf{$10^3$ users} & \textbf{$10^4$ users} & \textbf{$10^5$ users} & \textbf{$10^6$ users}  \\
\midrule
FoldFlow  & 0.970 & 0.970 & 0.943 & 0.900   \\
FrameDiff  & 0.705 & 0.393 & 0.309 & 0.225   \\
FrameFlow  & 1.000 & 0.992 & 0.980 & 0.931   \\
MultiFlow  & 1.000 & 0.996 & 0.940 & 0.817 \\
ESMFold & 0.903 & 0.824 & 0.450 & 0.334 \\
\bottomrule
\end{tabular}
\end{table*}


     



     


\subsection*{Robustness against post-processing and adaptive attacks}
In real applications, the malicious user may take post-processing or design adaptive attacks to bypass the safeguarding of FoldMark. Here, we consider three common post-processing methods for the protein structure and two adaptive attacks in Table. \ref{tab:adaptive}. Adaptive attacks involve fine-tuning the watermarked model using clean protein data to erase the watermark, or performing a multi-message attack, where additional watermarks are injected to obscure the original ones.
We can observe that FoldMark is robust to cropping, translation, and rotation because the watermark information is encoded into each residue and the watermark decoder is SE(3) invariant. 
Due to the integrated design and data augmentation, FoldMark is resistant to finetuning and multi-message injection.

\section*{Discussion}
In this paper, our study demonstrates the feasibility of embedding watermarks into protein generative models and their outputs through our proposed method, FoldMark. This two-stage approach successfully preserves the quality of protein structures while embedding user-specific information for copyright authentication and tracking. Extensive experiments on various protein structure prediction and design models confirm the effectiveness and robustness of FoldMark against post-processing and adaptive attacks, with minimal impact on the original structure quality. This provides a potential solution for addressing ethical concerns, such as copyright protection, in the application of generative AI to protein design.

There are a few limitations that we would like to address in the future. First, our approach struggles with significant structural modifications such as large-scale domain movements or extreme conformational changes, as the watermark's resilience is limited. Currently, the watermark pretraining process is decoupled from the fine-tuning of protein generative models, and future improvements in building end-to-end watermark pipelines could enhance robustness against such structural changes.
Additionally, advanced users may apply protein generative models not only for de novo design but also for structure editing, functional optimization, or motif scaffolding. At present, our watermarking technique does not sufficiently address these types of complex modifications, limiting its effectiveness in more advanced usage scenarios.
Finally, as the complexity or length of the generated protein increases, we observe some performance degradation in watermark retrieval accuracy. We plan to address this limitation in future work by optimizing our method to scale effectively with larger and more intricate protein structures.

\section*{Methods}
Figures \ref{pretrain} and \ref{finetune} provide an overview of our method. Inspired by previous works \cite{fernandez2023stable, feng2024aqualora, min2024watermark}, FoldMark consists of two main stages: Watermark Encoder/Decoder Pretraining and Consistency-Preserving Finetuning. The pretraining stage enables the watermark encoder and decoder to learn how to embed watermark information into the structure space and accurately extract it. The finetuning stage equips pretrained protein generative models with watermarking capabilities while preserving their original generative performance (Consistency-preserving). FoldMark is a versatile method that can be applied to various mainstream protein structure generative models. We use a diffusion-based model as an example and present the details of FoldMark below.

\subsection*{Watermark Encoder/Decoder Pretraining}
We first train a watermark encoder $\mathcal{W}$ and decoder $\mathcal{D}$ such that $\mathcal{D}$ can correctly retrieve the watermark message $\mathbf{m}$ embedded by $\mathcal{W}$.
\begin{equation}
    \mathcal{L}_{Pretrain} = \mathbb{E}_{\mathbf{x}, \mathbf{m},f}[\mathcal{L}_{BCE}(\mathcal{D}(f(\mathcal{W}(\mathbf{x}, \mathbf{m}))),\mathbf{m}) + \gamma\|\mathcal{W}(\mathbf{x}, \mathbf{m})-\mathbf{x}\|_2],
\end{equation}
where $\mathbf{x}$ represents the protein structure data and $\mathbf{m}$ denotes the string of binary watermark code.
$\gamma > 0$ is a hyperparameter to control the strength of structure adjustment for watermarking. $f$ represents a randomly selected structure distortion as data augmentation. The pool of data augmentation includes random rotation/translation, adding Gaussian noise to protein coordinates, and randomly cropping the protein structure. $\mathcal{L}_{BCE}(\mathcal{D}(f(\mathcal{W}(\mathbf{x}, \mathbf{m}))),\mathbf{m})$ and $\|\mathcal{W}(\mathbf{x}, \mathbf{m})-\mathbf{x}\|_2$ correspond to the CE Loss and Struct Loss in Figure \ref{pretrain} respectively.

\subsection*{Consistency-preserving Finetuning}

Instead of finetuning all the parameters of the generative model, we selectively fine-tune part of the protein generative model with WaterLoRA and the watermark decoder as shown in Figure \ref{finetune}. The other parameters including the watermark embedder $\mathcal{P}$ and the reference model are kept unchanged. We discuss the details of watermark module in the next subsection.

Here we take the diffusion-based protein generative model (e.g., FrameDiff \cite{yim2023se} and RFDiffusion \cite{watson2023novo}) as an example to construct the fine-tuning loss. The diffusion model typically involves two critical components known as the forward and backward process, where the forward process gradually noises the original protein structure $\mathbf{x}_0$ into $\mathbf{x}_t$ for $t\in\{1,\cdots, T\}$ and the model learns to predict the original structure $\epsilon_{\theta}(\mathbf{x}_t)$ based on $\mathbf{x}_t$. There are two losses in the fine-tuning: the consistency loss for regularization and the message retrieval loss to encourage correct watermark retrieval. In the consistency loss $\mathcal{L}_c$, the prediction of the fine-tuned model is compared with the original pretrained model so that the finetuned model weights will not deviate too much from the original ones. For the watermark retrieval loss $\mathcal{L}_{m}$, we take a single reverse step with respect to $\mathbf{x}_t$ to obtain $\hat{\mathbf{x}}_t = (\mathbf{x}_t - \sqrt{1-\Bar{\alpha_t}}\epsilon_{\theta}(\mathbf{x}_t))/ \sqrt{\Bar{\alpha_t}}$, and then feed it into the decoder to predict the watermark code.
In sum, we incorporate both optimization objectives above and formulate the consistency-preserving finetuning loss as:
\begin{equation}
\begin{aligned}
    \mathcal{L}_{Finetune} = \mathbb{E}_{\mathbf{x}, t, \mathbf{m}}[\mathcal{L}_c(\epsilon_{\theta}(\hat{\mathbf{x}}_{t}), \epsilon_{\theta_{ref}}(\mathbf{x}_t)) + \eta\cdot\frac{t-T}{T}\mathcal{L}_{m}(\mathcal{D}(\hat{\mathbf{x}}_t), \mathbf{m})],
\end{aligned}
\end{equation}
where $\eta$ controls the trade-off between consistency loss $\mathcal{L}_c$ and watermark retrieval loss $\mathcal{L}_{m}$. We place an additional weight $\frac{t-T}{T}$ for the retrieval loss because the generated structure contains more information of watermark as $t\rightarrow 0$ and we observe better performance in experiments.  

\subsection*{Watermark-conditioned LoRA (WaterLora)}
To save the computation costs of fine-tuning and flexibly embed watermark information in the generation process, we propose the Watermark-conditioned LoRA (WaterLoRA) strategy, inspired by previous works \cite{hu2021lora, kopiczko2023vera, zhang2023adalora, feng2024aqualora, si2024flora} on parameter-efficient fine-tuning and modulation. The computation formula for WaterLoRA can be expressed as:
\[
\Delta \mathbf{W}(\mathbf{m}) = \mathbf{G}(\mathbf{m}) \odot (\mathbf{A} \times \mathbf{B}),
\]
where \( \mathbf{A} \in \mathbb{R}^{n \times r} \) and \( \mathbf{B} \in \mathbb{R}^{r \times m} \) are the low-rank matrices, and \( \mathbf{G} \in \mathbb{R}^n \) is the gating vector derived from the watermark code. The operator \( \odot \) denotes element-wise multiplication, where \( \mathbf{G} \) modulates the rows of \( \mathbf{A} \times \mathbf{B} \). This formulation maintains efficiency while allowing flexible incorporation of watermark information.

To input the watermark information into the fine-tuned model, we utilize an adapter layer that converts a watermark code of length \( l \) into a gating vector \( \mathbf{G} \). Specifically, the watermark code \( \mathbf{m} = \{b_0, b_1, \ldots, b_l\} \) is passed through a linear transformation defined as:
\[
\mathbf{G}(\mathbf{m}) = \mathbf{W}_g \cdot \mathbf{m} + \mathbf{b}_g,
\]
where \( \mathbf{W}_g \in \mathbb{R}^{n \times l} \) is the linear transformation matrix, and \( \mathbf{b}_g \in \mathbb{R}^n \) is the bias vector. Here, \( b_i \in \{0, 1\} \) represents the binary state of the \( i \)-th bit in the watermark code. 
The gating vector \( \mathbf{G} \) modulates the LoRA weight updates by scaling the rows of the low-rank update \( \mathbf{A} \times \mathbf{B} \).

During the generation process, when embedding a watermark into the model, we compute the gating vector \( \mathbf{G} \) based on the watermark code. The resulting LoRA weight update \( \Delta \mathbf{W} \) is added to the original model weights to produce the watermarked model weights:
\[
\mathbf{W}_{\text{watermarked}} = \mathbf{W} + \alpha \Delta \mathbf{W} (\mathbf{m}),
\]
where \( \alpha \) is a scaling factor controlling the impact of the watermark on the model weights. FoldMark apply WaterLora to all linear and attention layers for effective watermarking.

\subsection*{Experimental Settings}
\paragraph{Datasets.} We trained watermark encoders/decoders and fine-tuned protein generative models using the monomers from the PDB \cite{berman2000protein} dataset, focusing on proteins ranging in length from 60 to 512 residues with a resolution better than 5 Å. This initial dataset consisted of 23,913 proteins. Following previous work \cite{yim2023se}, we refined data by applying an additional filter to include only proteins with high secondary structure content. For each monomer, we used DSSP \cite{kabsch1983dictionary} to analyze secondary structures, excluding those with over 50\% loops. This filtering process resulted in 20,312 proteins. 

\paragraph{Implementations.}
Our FoldMark model is pretrained for 20 epochs and fine-tuned for 10 epochs with Adam \cite{kingma2014adam} optimizer, where the learning rate is 0.0001, and the max batch size is 64. 
We use the batching strategy from FrameDiff \cite{yim2024improved} of combining proteins with the same length into the same batch to remove extraneous padding. In the WaterLoRA, the rank is set as 16 in the default setting. $\gamma$ and $\eta$ are set as 2.
We report the results corresponding to the checkpoint with the best validation loss. It takes less than 48 hours to finish the whole training process on 1 Tesla A100 GPU. More hyperparameter settings are listed in Table. \ref{tab:hyperparam}.

\paragraph{Baselines.} To the best of our knowledge, FoldMark is the first watermarking method specifically designed for protein structure generative models. For comparison, we adapted two state-of-the-art watermarking methods originally developed for image generation: WaDiff \cite{min2024watermark} \footnote{https://github.com/rmin2000/WaDiff} and AquaLoRA \cite{feng2024aqualora} \footnote{https://github.com/Georgefwt/AquaLoRA}. Both baseline models were designed for image diffusion models, such as Stable Diffusion \cite{rombach2022high}. Since most protein generative models are also diffusion-based, we applied the recommended hyperparameters from the original works.

\section*{Data availability}
This study's training and test data are available at Zenodo (\url{https://github.com/zaixizhang/FoldMark}). The project website for FoldMark is at \url{http://home.ustc.edu.cn/~zaixi/projects/FoldMark}.

\section*{Code availability}
The source code of this study is freely available at GitHub (\url{https://github.com/zaixizhang/FoldMark}) to allow for replication of the results of this study.

\bibliography{main}

\clearpage

\section*{Author contributions statement}
Z.X.Z., M.Z., and M.D.W. designed the research,  Z.X.Z. and K.D.F. conducted the experiments, Z.X.Z., L.C, M.Z., and M.D.W. analyzed the results. Z.X.Z., R.F.J., L.C., M.Z., and M.D.W. wrote the manuscript.  All authors reviewed the manuscript. 

\section*{Competing interests}
The authors declare no competing interests.

\section*{Additional information}

{\bf Correspondence and requests for materials} should be addressed to Mengdi Wang.

\clearpage
\setcounter{figure}{0}

\clearpage
\setcounter{figure}{0}
\setcounter{table}{0}
\makeatletter 
\renewcommand{\thefigure}{S\@arabic\c@figure}
\renewcommand{\thetable}{S\@arabic\c@table}
\makeatother

\section*{Supplementary Information}
\subsection{Comparison with other watermark methods}
Traditional watermarking techniques developed for Large Language Models (LLMs) and diffusion models are not directly applicable to protein structure data due to the unique characteristics of protein structures. Similar methods such as WaDiff \cite{min2024watermark} and AquaLoRA \cite{feng2024aqualora}, which embed watermarks into the U-Net backbone of Stable Diffusion models for image generation, are effective in the image domain but typically fail in protein generative models. FoldMark addresses this gap by employing optimized pretraining and fine-tuning losses, a customized encoder-decoder architecture, and enhanced watermark-conditioned Low-Rank Adaptation (LoRA) methods, achieving superior performance in protecting protein generative models.

\subsection{Ablation studies of FoldMark}

\subsection{Watermark Encoder-Decoder architecture}

\end{document}